# Theoretical studies of the kinetics of mechanical unfolding of cross-linked polymer chains and their implications for single molecule pulling experiments

**Kilho Eom**[§]**, Dmitrii E. Makarov**[¶†‡]**, and Gregory J. Rodin**[§†]

*Department of Aerospace Engineering & Engineering Mechanics*[§], *Department of Chemistry & Biochemistry*[¶], *Institute for Computational Engineering & Science*[†], *and Institute for Theoretical Chemistry*[‡], *The University of Texas at Austin, Texas 78712*

**Abstract**

We have used kinetic Monte Carlo simulations to study the kinetics of unfolding of cross-linked polymer chains under mechanical loading. As the ends of a chain are pulled apart, the force transmitted by each crosslink increases until it ruptures. The stochastic crosslink rupture process is assumed to be governed by first order kinetics with a rate that depends exponentially on the transmitted force. We have performed random searches to identify optimal crosslink configurations whose unfolding requires a large applied force (measure of strength) and/or large dissipated energy (measure of toughness). We found that such optimal chains always involve cross-links arranged to form parallel strands. The location of those optimal strands generally depends on the loading rate. Optimal chains with a small number of cross-links were found to be almost as strong and tough as optimal chains with a large number of cross-links. Furthermore, optimality of chains with a small number of cross-links can be easily destroyed by adding cross-links at random. The present findings are relevant for the interpretation of single molecule force probe spectroscopy studies of the mechanical unfolding of "load-bearing" proteins, whose native topology often involves parallel strand arrangements similar to the optimal configurations identified in the study.

# 1. Introduction

A number of proteins exhibit a combination of strength and toughness that cannot be matched by artificial materials[1-4]. Recent single molecule force probe spectroscopy experiments suggest that these remarkable properties are accomplished through the mechanical response of individual protein domains, which are capable of dissipating large energy upon their mechanical unfolding[2,4,5].

In single molecule pulling experiments employing the atomic force microscope (AFM), one end of the protein is attached to a substrate and the other end is attached to a cantilever(see, e.g., refs. [6-8] for a review); The cantilever then can be displaced at a constant rate. During such an experiment, one measures the pulling force, and then presents the data in the form of the force-displacement curve. The forces generated by different proteins under typical experimental conditions range from a few piconewtons to several hundred piconewtons and generally depend on the pulling rate. If one were to perform an equilibrium, reversible stretching experiment by pulling on the molecule at a sufficiently slow rate then the measured force-vs-displacement curve would become rate independent and the work done by the pulling force would be equal to the free energy difference between the folded and the stretched states of the molecule. In practice, stretching of a molecule is nearly an equilibrium process if the time scale of pulling is longer than that of the molecule's conformational changes. This equilibrium regime is rarely achieved in AFM pulling studies. It further appears that many proteins that perform "load-bearing" functions in living organisms operate far away from equilibrium; as a result their mechanical stability is often uncorrelated with their thermodynamic stability[7,9-12].

For example, the work required to unfold the molecule of the muscle protein titin in a typical AFM pulling experiment is about two orders of magnitude higher than its free energy of folding, indicating that this is a highly nonequilibrium process[5]. This property accounts for the high toughness of titin arguably required for its biological function in the muscles. Similarly, the difference between the force-vs.-extension curves measured in the course of stretching and subsequent relaxation of spider capture silk proteins[4] reveals that stretching is a non-equilibrium process, in which extra energy is dissipated. In contrast, the work required to unfold of the myosin coiled-coil via pulling on it at similar pulling rates is comparable to the free energy of folding, indicating that this is a nearly equilibrium process[5].

The mechanical resistance of a protein is thus determined both by its structure and by the loading rate. Recently, we have studied a toy model of a cross-linked polymer chain, which we used to identify the chain configurations that lead to its high mechanical resistance[13]. In that model, we considered a Gaussian chain with rigid cross-links. Unfolding of the chain under mechanical loading occurs as a result of rupture of the cross-links. Each cross-link ruptures once its internal force reaches a critical value. Thus, as the chain ends are being pulled apart at a constant rate, the force in each link increases until it ruptures. As the loading proceeds, all the crosslinks become ruptured and the chain unfolds. The excess work done on the cross-linked chain, as compared to the work done stretching the unconstrained chain, is a measure of the chain toughness. Given the total number of cross-links, one may seek the optimal cross-link configurations that maximize either the excess work or the maximum force during the unfolding process. Our rationale for studying such a simple model was the previous finding[7,10-12,14] that the unfolding mechanism is largely determined by the native topology of the protein. This view is further supported by the success of simplified, Go-like models in predicting the mechanisms of mechanical unfolding[15-18]. Although Gaussian cross-linked chains are merely caricatures of real biopolymers, they may adequately capture the effects of topology on the unfolding mechanism. Indeed, there are good reasons to believe they do. Specifically, the key finding of our previous study is that the optimal configurations that maximize the peak force and the dissipated energy must involve parallel strands. This finding is consistent with experimental studies[7,9,10,19-24] and molecular dynamics simulations[25-29] of the protein domains exhibiting high unfolding forces, such as the I27 domain in titin. Further, this finding has led to the prediction that protein domains with the ubiquitin fold, which features terminal parallel strands similar to those in I27, exhibit superior mechanical properties, despite the fact that they have no apparent mechanical functions in living organisms[30]. This prediction is supported by both experiments[12] and molecular dynamics simulations[30,31].

While providing results that are qualitatively consistent with atomistic scale studies, our model[13] entirely ignored stochastic and rate-dependent aspects of unfolding.

This is an unrealistic assumption in many cases because, in general, rupture of a chemical bond is a chemical reaction, i.e., a stochastic process whose rate is affected by the transmitted force[32]. Further, as we mentioned earlier, load-bearing proteins exhibit high toughness and strength precisely because they are loaded at high rates so that unfolding is a non-equilibrium irreversible process accompanied by large energy dissipation.

Models of force-induced rupture of chemical bonds are well known in the contexts of protein unfolding and ligand unbinding[19,20,32-35] and fracture[36]. In those models, rupture of a bond is described by first order kinetics and its rate depends on the force transmitted by the bond. The main purpose of this paper is to adapt our model of cross-linked Gaussian chains to study how the optimal chain configurations that maximize the excess work and/or the maximum force depend on the loading rate. To this end, we have assumed that rupture of each cross-link is described by first order kinetics with a force-dependent probability and performed kinetic Monte Carlo studies of the chain unfolding. The main finding of this study is that the parallel-strand arrangements remain optimal even when the stochastic nature of bond breaking is taken into account; While always featuring such parallel strands, the found optimal configurations generally depend on the loading rate.

The rest of this paper is organized as follows. In Section 2, we describe the model. In Section 3, we outline the simulation methods. In Section 4, we present our simulation results. In Section 5, we discuss implications of our results for pulling experiments on single molecules.

## 2. The model

Consider a polymer chain consisting of $L+1$ beads connected by $L$ links. The chain is assumed to obey Gaussian statistics so that the probability distribution for the distance between beads $i$ and $j$ is given by

$$P(|\mathbf{r}_i - \mathbf{r}_j|) = \left[\frac{3}{2\pi b^2 |i-j|}\right]^{3/2} \exp\left[-\frac{3|\mathbf{r}_i - \mathbf{r}_j|^2}{2|i-j|b^2}\right], \qquad (1)$$

where $b$ is the rms length of a single link. One way to construct such a Gaussian chain is to connect neighboring beads by harmonic springs so that its potential energy is given by

$$U = \frac{1}{2} g_0 \sum_{i=1}^{L} |\mathbf{r}_{i+1} - \mathbf{r}_i|^2 \quad \text{with} \quad g_0 = \frac{3k_B T}{b^2} \qquad (2)$$

where $k_B$ is Boltzmann's constant and $T$ is the temperature.

The motion of the chain is constrained by $N$ cross-links. Each link is designated by the indices of its end points, so that the entire set of cross-links is denoted by $C_N = \{\{i_1, j_1\}, \cdots, \{i_N, j_N\}\}$. Each cross-link is regarded as rigid; alternatively, one can model a cross-link as a spring whose stiffness $g_c \gg g_0$. We assume that no bead can be attached to more than one cross-link, so that the maximum number of cross-links is $N = (L+1)/2$.

The chain ends (monomers number 1 and $L+1$) are pulled apart at a constant speed $v$ so that the distance between them grows linearly as a function of time $t$:

$$|\mathbf{r}_L - \mathbf{r}_0| \equiv e = vt \tag{3}$$

We suppose that loading is slow compared to a typical time scale of thermal Brownian motion of the chain. In this case, we assume that the value of the pulling force $F(t)$ recorded at any instant $t$ is the force *averaged* over the thermal motion. At the same time, the time scale of cross-link rupture may be comparable with that of loading and so the rupture of a cross-link may result in a measurable change in $F(t)$.

We consider two rupture models for the cross-links. In the first model, to which we refer as *Model I*, a cross-link ruptures deterministically once its internal force reaches a critical value $f_c$. This model has been studied previously[13] but we include it here for comparisons. In the second model, to which we refer as *Model II*, rupture of a cross-link is a stochastic process described by first-order kinetics. Specifically, the probability that the cross-link ruptures in the time interval from $t$ to $t+\Delta t$ depends only on the instantaneous value of the internal force $f(t)$ and is given by[32]

$$k[f(t)]\Delta t = k_0 \exp\left[\frac{f(t)}{f_c}\right]\Delta t, \tag{4}$$

where $k_0$ is the rupture rate constant at zero force and $f_c$ is a reference force. Equation (4) is a commonly used model, which assumes that the free energy barrier to rupture decreases linearly with the force $f$[20,32]. Although this equation is not necessarily quantitative[31,37], it is sufficient for qualitative predictions, as it properly identifies the rapid increase in the probability of rupture once the internal force exceeds $f_c$.

Because the rate of Eq. 4 is not zero at zero force, then, strictly speaking, any cross-link configuration in Model II is unstable and the chain will unfold irreversibly on a time scale of order $k_0^{-1}$ even if no force is applied. This is not realistic since the folded state of a protein at zero force is expected to be thermodynamically more stable than its unfolded state. It is necessary to allow for the recombination of cross-links in order to restore the detailed balance in the system[35,38,39]. Here, we assume that the time $k_0^{-1}$ is much longer than the time scale of loading. Under this assumption recombination of cross-links during unfolding is unlikely because forces in each crosslink will quickly reach values large enough to destabilize each bond thermodynamically such that the ruptured bond state has lower free energy than that with the bond intact; In other words, once the bond is broken it will be unlikely to reform unless the loading force is removed. For these reasons we did not include cross-link recombination in our model; It would therefore not be applicable to very slow, nearly equilibrium pulling experiments. In this respect, the physical regime explored by the present work is quite different from the reversible stretching conditions assumed in the theoretical studies of RNA and DNA mechanical denaturation[40-44] and in the theories of the reversible stretching of protein-like heteropolymers[45-47]. Note, however, that nonequilibrium effects have been considered in ref.[46]

When the ends of a Gaussian chain are pulled apart, its response follows Hooke's law[48,49], which also holds in the presence of cross-links[50]. However the stiffness of the chain changes upon cross-link rupture. Under constant velocity loading conditions, the force-displacement curve $F = F(e)$ is a peace-wise linear function with jumps and different slopes (see Fig. 1). Once all the cross-links are ruptured, the slope is reduced to the stiffness of the unconstrained chain, $\Gamma_0 = g_0 / L$.

The mechanical response of a cross-linked chain is represented by two quantities (cf. Fig. 1): (*i*) the maximum force $F_m$ and (ii) "toughness", i.e., the excess work done upon unfolding:

$$\Delta W = \int_0^u F(e)de - \frac{1}{2}\Gamma_0 u^2 \qquad (5)$$

where $u$ is the distance between the $1^{st}$ and the $L+1^{st}$ beads at the end of the pulling experiment, once all the cross-links have been ruptured.

For Model I, rupture is a deterministic process, so that $F_m$ and $\Delta W$ are unique for a given set $C_N$. Further, the force-displacement curve and its parameters $F_m$ and $\Delta W$ can be determined upon solving a set of $N$ linear problems that reflect the sequence of the rupture events. In contrast, in Model II, rupture is a stochastic process. Accordingly, for a given set $C_N$, it is necessary to determine the averages of $F_m$ and $\Delta W$ over sufficiently large number of realizations of the stochastic unfolding process; we denote those quantities by $\langle F_m \rangle$ and $\langle \Delta W \rangle$, respectively.

The adopted model will be used in the following settings:

- *Characterization problem*: Given $L$, $C_N$, $g_0$, $k_0$, $f_c$, and $v$ determine $\langle F_m \rangle$ and $\langle \Delta W \rangle$.

- *Optimization problem*: Given $L$, $N$, $g_0$, $k_0$, $f_c$, and $v$ determine the

configuration(s) $C_N$ that maximize(s) $\langle F_m \rangle$ and $\langle \Delta W \rangle$.

## 3. Methods

### 3.1. Elasticity analysis

Between two rupture events, the cross-linked chain responds as a collection of Hookean springs [50]. The springs are identified as follows:

1. Arrange the $2N$ beads belonging to the cross-links in the ascending order:

$$1 \leq i_1 < i_2 < ... i_{2N-1} < i_{2N} \leq L$$

2. Identify each chain segment between two consecutive members of this set as a spring.

3. Assign to each springs the stiffness $\gamma_0/n$, where $n$ is the number of the chain links in the segment.

Once the springs and their stiffness have been identified, the entire assembly can be analyzed using the finite element method [50]. The results can be expressed as

$$F(t) = \Gamma(t)vt \tag{6}$$

and

$$f_k(t) = \mathbf{a}_k(t) F(t), \tag{7}$$

where $\Gamma(t)$ is the instantaneous overall stiffness of the cross-linked chain, $f_k(t)$ is the internal force in the $k$-th crosslink, and $\alpha_k(t)$'s are dimensionless coefficients. Note that $\Gamma(t)$ and $\alpha_k(t)$ depend on the current configuration of the cross-links and remain constant between rupture events; in general, they are piecewise constant functions of time.

## 3.2. Kinetic Monte Carlo method

To simulate the stochastic unfolding process we use the kinetic Monte Carlo method[35,51-53]. Suppose that at time $t_0$, there are $n$ cross-links. Let us evaluate the probability that the first rupture among those cross-links occurs at a later time, in the time interval between $t$ and $t+\Delta t$. This probability is equal to the probability $S(t,t_0)$ that no cross-link has ruptured in the time interval between $t_0$ and $t$, times the sum of the probabilities for each of the cross-link to rupture in the time interval between $t$ and $t+\Delta t$:

$$\Phi(t)\Delta t = S(t,t_0)\sum_{m=1}^{n} k[f_m(t)]\Delta t \tag{8}$$

Also, in the time interval between $t$ and $t+\Delta t$ the survival probability is reduced by $\Phi(t)\Delta t$, so that

$$-\Phi(t)\Delta t = S(t+\Delta t,t_0) - S(t,t_0) \tag{9}$$

By combining equations 4, 6-9 we obtain

$$S(t,t_0) = \exp\left\{-k_0\sum_{m=1}^{n}\frac{f_c}{\mathbf{a}_m(t_0)\Gamma(t_0)v}\left[\exp\left(\frac{\mathbf{a}_m(t_0)\Gamma(t_0)vt}{f_c}\right) - \exp\left(\frac{\mathbf{a}_m(t_0)\Gamma(t_0)vt_0}{f_c}\right)\right]\right\} \tag{10}$$

and

$$\Phi(t) = k_0 S(t,t_0)\sum_{m=1}^{n}\exp\left[\frac{\mathbf{a}_m(t_0)\Gamma(t_0)vt}{f_c}\right]. \tag{11}$$

The standard method[35,51-53] for generating the time $t$ on a computer is to solve the equation

$$S(t,t_0) = \mathbf{x} \tag{12}$$

where $\mathbf{x}$ is a uniformly distributed random variable in the interval [0,1]. We use modified Newton's method to solve this equation numerically. Once the time $t$ is generated, we need to determine which of the $n$ cross-links ruptures. This is done by computing the weighted probability of rupture for each of the cross-links:

$$w_m = \frac{\exp\left[\frac{f_m(t)}{f_c}\right]}{\sum_{j=1}^{n}\exp\left[\frac{f_j(t)}{f_c}\right]} \quad \text{with} \quad m=1,\ldots,n \tag{13}$$

Next, we divide the interval [0,1] into $n$ sub-intervals whose lengths are $w_m$. Finally, we generate $\lambda$, a realization of a random variable uniformly distributed in the interval [0,1], and identify the sub-interval containing $\lambda$. The index of this sub-interval is equal to the index of the cross-link to be ruptured This process is followed starting with $t=0$, $n=N$ and until all the cross-links are ruptured.

The quantities $\langle F_m \rangle$ and $\langle \Delta W \rangle$ for a given set $C_N$ are computed by averaging over $N_{MC}$ realizations of the unfolding history; we used $N_{MC} = 5,000$.

**3.3 Optimization**

We used two optimization methods for finding the configurations that maximize $\langle F_m \rangle$ and/or $\langle \Delta W \rangle$. In cases where the search space was sufficiently small, we exhaustively searched over all possible sets $C_N$. When an exhaustive search was too time-consuming, we resorted to the following "random hill-climbing" procedure[13]:

(1) Generate a random set $C_N^{(0)}$ with $N$ cross-links.

(2) Select a cross-link $\{i, j\}$ from the set $C_N^{(0)}$.

(3) Evaluate $\langle F_m \rangle$ (or $\langle \Delta W \rangle$) for $C_N^{(0)}$, and the "adjacent" sets obtained from $C_N^{(0)}$ upon replacing $\{i,j\}$ with $\{i, j\pm 1\}$ or $\{i\pm 1, j\}$. Of course, the sets $\{i, j\pm 1\}$ and $\{i\pm 1, j\}$ must be admissible, in the sense that no bead can be connected to more than one cross-link.

(4) Choose the optimal set among the five sets identified at Step 3.

(5) Repeat Steps 2-4 for all other cross-links to complete the first sweep. This defines a new configuration $C_N^{(1)}$

(6) Repeat Steps 1-5 until $C_N^{(i+1)} = C_N^{(i)}$

(7) Generate new $C_N^{(0)}$ and repeat steps 2-6.

## 4. Results

### 4.1 Single cross-link

*Model I.* A single cross-link, $\{i, i+l\}$, creates a loop of length $l$ in the chain. The optimal configurations in this case can be found analytically[13]. In particular, $F_m = f_c$ for all $i$ and $l$, and $\langle \Delta W \rangle$ depends on $l$ only:

$$\Delta W = \frac{f_c^2}{2\Gamma_0}\left(\tilde{l} - \tilde{l}^2\right), \tag{16}$$

where we have introduced the dimensionless loop length

$$\tilde{l} = \frac{l}{L}$$

The excess work reaches its maximum for $\tilde{l} = 1/2$:

$$\Delta W = \frac{f_c^2}{8\Gamma_0}.$$

Thus one can regard the configurations with $\tilde{l} = 1/2$ as optimal with respect to both $F_m$ and $\langle \Delta W \rangle$.

*Model II*. The model parameters give rise to the dimensionless time

$$t = k_0 t$$

and dimensionless pulling rate

$$\tilde{v} = \frac{\Gamma_0 v}{k_0 f_c}.$$

Following the analysis in Section 3.2, it is straightforward to obtain the probability density function for the dimensionless rupture time $t$,

$$\Phi(t, q) = \exp(qt) \exp\left\{\frac{1}{q}\left[1 - \exp(qt)\right]\right\}, \tag{17}$$

where the parameter $q$ combines the dimensionless loading rate and geometric parameters,

$$q = \frac{\tilde{v}}{1 - \tilde{l}}.$$

This combination arises naturally for *N*=1 but not for *N*>1. At the moment of rupture we have

$$F_m(t) = f(t) = \frac{\Gamma_0}{1 - \tilde{l}} v k_0 t = f_c q t \tag{18}$$

and

$$\Delta W(t) = \frac{1}{2}\Gamma_0 (vk_0 t)^2 \left(\frac{1}{1-\tilde{l}} - 1\right) = \frac{1}{2}\frac{f_c^2}{\Gamma_0}(1-\tilde{l})\tilde{l}q^2 t^2,$$

and therefore we obtain

$$\langle F_m \rangle = f_c q \int_0^\infty t \Phi(t,q) dt$$

and

$$\langle \Delta W \rangle = \frac{1}{2}\frac{f_c^2}{\Gamma_0}(1-\tilde{l})\tilde{l}q^2 \int_0^\infty t^2 \Phi(t,q) dt$$

The integrals involved in these expressions can be evaluated numerically only. Nevertheless, one can obtain asymptotic approximations valid for $q \gg 1$:

$$\langle F_m \rangle \approx f_c \ln q = f_c \ln \frac{\tilde{v}}{1-\tilde{l}} \tag{19a}$$

$$\langle \Delta W \rangle = \frac{1}{2}\frac{f_c^2}{\Gamma_0}(\tilde{l}-\tilde{l}^2)\ln^2 q = \frac{1}{2}\frac{f_c^2}{\Gamma_0}(\tilde{l}-\tilde{l}^2)\ln^2 \frac{\tilde{v}}{1-\tilde{l}}. \tag{19b}$$

The meaning of Eq. 19a is simple: This is the force (Eq. 18) corresponding to the most probable rupture time that maximizes the probability density of Eq. 17. As expected, this asymptotic expression for $\langle F_m \rangle$ reveals the logarithmic dependence on the loading rate[32-34]. Further, $\langle F_m \rangle$ increases indefinitely as $\tilde{l} \to 1$, i.e. the largest forces are generated by chains with terminal cross-links. The case of $\tilde{l}=1$ corresponds to a chain whose ends cannot be pulled apart, and therefore it is meaningless in the context of the present model.

The excess work also grows logarithmically with $\tilde{v}$, but in contrast to $\langle F_m \rangle$, its optimization leads to values of $\tilde{l}$ that depend on $\tilde{v}$. In particular, for $\tilde{v} \to \infty$ the optimal value is $\tilde{l} \to 1/2$. In general, for moderately large values of $\tilde{v}$ the optimal value of $\tilde{l}$ is in the range $1/2 < \tilde{l} < 1$ (see Table 1). All of these conclusions are straightforward to derive from the asymptotic approximations of Eq. 19 and are confirmed by computing the exact expressions.

It is instructive that the optimal chain configuration maximizing the excess work $\langle \Delta W \rangle$ in Model II in the limit of infinitely fast loading is the same as the optimal configuration predicted by Model I. The fast pulling limit of Model II, where a crosslink rupture is unlikely until the internal force attains a sufficiently large value, $f \geq f_c$, can be roughly approximated by Model I. The two models however do not become equivalent in this limit: The unfolding force for a single crosslink is independent of the chain configuration and equal to a constant value of $f_c$ in Model I while it depends on both on the loading rate and the crosslink location in Model II.

**4.2 Small number of cross-links**

*Model I.* This case has been studied in detail in ref. [13] The key result is that the same optimal configurations maximize both $F_m$ and $\Delta W$. Those configurations involve "parallel strands" of the form $C_N = \{\{i_1, j_1\}, \{i_2, j_2\}, \ldots \{i_N, j_N\}\}$ such that $i_1 < i_2 < \ldots i_N < j_1 < j_2 < \ldots < j_N$. For example, for $N=3$ and $L=50$ the optimal configurations have the form $\{\{i,i+l\},\{i+1,i+l+2\},\{i+3,i+l+3\}\}$ where $l=26$ (see Fig. 2). Note that the optimal value of $l$ is $l \approx L/2$, which is similar to that found in the case of a single crosslink.

Further, we showed that optimality can be understood in terms of a continuous "super cross-link" (SCL) model. In the limit as the chain becomes continuous, that is $L \to \infty$ and $b \to 0$, the topological constraint that any bead can be connected to only one cross-link can be relaxed because, as far as the mechanical response is concerned, neighboring beads become indistinguishable. Therefore one can create a SCL by placing all $N$ cross-links between the same points, $\{i, i+l\}$. Then the cross-links share the load equally so that the force in each cross-link is $F/N$, and the SCL acts like a single cross-link that can sustain a maximum force of $F_m = Nf_c$ resulting in an excess work of unfolding equal to (cf. Eq. 16):

$$\Delta W = \frac{N^2 f_c^2}{2\Gamma_0}\left(\tilde{l} - \tilde{l}^2\right)$$

As in the case of $N=1$, the maximum $\Delta W$ is achieved when $\tilde{l} = 1/2$.

For a discrete chain, we cannot achieve the SCL configurations because of the imposed constraint prohibiting multiple cross-links between the same monomers. Nevertheless, it turns out that the constrained optimal solutions are very close to the SCL's, and they involve parallel strands. We refer to such configurations as "nearly super cross-links" or NSCL's (Fig. 2). The force in each of the cross-links in the NSCL configuration is approximately the same. Further, within Model I, rupture of one cross-link in an NSCL configuration results in an increase of the force in each of the remaining cross-links such that NSCL's rupture in an avalanche-like fashion. Because of that the force vs. displacement curve $F(e)$ has only a single maximum, similarly to the case of a single cross-link.

*Model II.* Remarkably, we found that the NSCL configurations appear to be optimal with respect to both $\langle F_m \rangle$ and $\langle \Delta W \rangle$, although the configurations optimal for $\langle F_m \rangle$ are not necessarily optimal for $\langle \Delta W \rangle$, and vice versa. This statement is difficult to verify conclusively, because even for $N=3$ the search space is too large for an exhaustive search. Nevertheless, using the search algorithm described in Section 3.3, we could not find a configuration better than the NSCL of the form $\{\{i,i+l\},\{i+1,i+2+l\},\{i+3,i+3+l\}\}$, where the optimal value of $l$ was determined by the exhaustive search with respect to $l$. The optimal values of $l$ maximizing $\langle F_m \rangle$ and $\langle \Delta W \rangle$ were different, which is similar to the conclusion reached with Model II for $N=1$. Furthermore, the values of $\tilde{l} = l/L$ that optimize $\langle F_m \rangle$ are close to $\tilde{l} = 1$ and the optimal values of $\tilde{l}$ that maximize $\langle \Delta W \rangle$ depend on $\tilde{v}$ in a way similar to the case of $N=1$ (See Table 2). We also found that $\langle F_m \rangle$ and $\langle \Delta W \rangle$ grow logarithmically with $\tilde{v}$ (Fig. 3).

An attempt to predict the response of NSCL configurations using the rate-dependent SCL model was only partially successful. In particular, the rate-dependent SCL model was able to follow the trends predicted by the simulations but the agreement was mostly qualitative. Furthermore, the predictions of the rate-dependent SCL model were qualitatively similar to those obtained from the analysis for *N*=1. Let us mention that the rate-dependent SCL model was successful in predicting the first but not the last rupture events, especially for intermediate loading rates. In the limit $\tilde{v} \to \infty$, one can use the asymptotic approximations developed for *N*=1, with the provision that $k_0$ and $f_c$ are replaced with $Nk_0$ and $Nf_c$, respectively.

### 4.3 Large number of cross-links

For $N \ll L$, $F_m$ and $\Delta W$ are proportional to $N$ and $N^2$, respectively. Preliminary computations [13] have suggested that these scaling rules do not hold for large *N*, as both $F_m$ and $\Delta W$ tend to saturate with increasing *N*.

Here we study in more detail the case where each bead is connected to another bead so that the total number of cross-links is $N = L/2$ (for an even *L*) or $(L+1)/2$ (for an odd *L*). In this case, the search space is large and for this reason we limited our analysis to short chains, $L = 19$, and to using Model I only. The key result of our computations can be stated as follows:

- All optimal configurations contained the subset of three cross-links

  $$C_3^* = \{\{i, i+L/2\}, \{i+1, i+3+L/2\}, \{i+4, i+4+L/2\}\},$$

  which, again, is a "clamp" of parallel strands. The excess work for the configuration $C_3^*$ in the absence of any other cross-links is equal to $\Delta W^* = 0.79 f_c^2 / \Gamma_0$.

- By adding seven random cross-links to the clamp one is more likely to reduce than

to increase $\Delta W$ in comparison to $\Delta W^*$.

- The maximum $\Delta W$ is $\Delta W_m = 0.93 f_c^2 / \Gamma_0$, corresponding to the configuration

  $C_m^{(10)} = \{\{1,15\},\{2,11\},\{3,16\},\{4,14\},\{5,17\},\{6,10\},\{7,9\},\{8,18\},\{13,20\},\{12,19\}\}$,

  which also maximizes $F_m$.

- The mean value of toughness for randomly generated cross-link configurations is $\Delta \bar{W} \approx 0.35 f_c^2 / \Gamma_0$, and only a small fraction of configurations have the toughness close to $\Delta W_m$.

These results are further illustrated in Figure 4, where we plot the histograms for $F_m$ and $\Delta W$ corresponding to randomly generated cross-link configurations and configurations containing the subset $C_3^*$. The latter, on the average, have larger values of both $F_m$ and $\Delta W$, as compared to random crosslink arrangements. However, adding random cross-links to $C_3^*$ does not necessarily improve the mechanical resistance of the chain: only a relatively small fraction of such configurations perform better than $C_3^*$.

## 5. Discussion: Implications for force-induced protein unfolding.

We have previously demonstrated[13] that a clamp formed by parallel strands represents the optimal topology maximizing both the unfolding excess work and the unfolding maximum force for the deterministic Model I, in which cross-links rupture once the transmitted force achieves a critical value. Molecular mechanics studies have also demonstrated that large forces are required to rupture parallel β-strands in proteins[29]. In the range of unfolding forces typical in force probe spectroscopy pulling experiments[7,8,19,21,24,54-60] the rupture of hydrogen bonds is a probabilistic phenomenon described by rate kinetics[32-35]. The present study demonstrates that parallel strands remain optimal even when such kinetic effects are important. Moreover, there is a close relationship between the fast pulling limit of the proposed kinetic model and the deterministic Model I, both predicting the same conformations optimal with respect to the excess work of unfolding.

Three findings of the present study may provide additional insight into the properties of strong proteins:

- For moderate pulling rates, parallel strands formed between the *ends of the chain* (i.e., those with $l \simeq L$) lead to higher values for both $F_m$ and $\Delta W$. In contrast, for very high pulling rates, parallel strands with $l \simeq L/2$ are optimal with respect to $\Delta W$. The former observation is consistent with both the experimental[9-12] and computational[25,26,30,31,37] evidence for (so far) known protein domains with superior mechanical properties – all of them contain *terminal* parallel β−strands.

- The configurations that include the optimal NSCL configurations are superior to random structures (see Section 4.3). This result may shed light on the recent finding that proteins with different folds may display similar mechanical resistance. In particular, the unfolding mechanisms of the all-β I27 domain of the muscle protein titin[19,20,25,26] and of the α/β ubiquitin domain[12,30] are very similar and are characterized by a high unfolding force because both of these domains feature the same hydrogen-bond clamp formed by their terminal parallel strands.

- Adding random crosslinks to an optimal NSCL configuration can be viewed to some extent as a way to mimic the effect of nonnative interactions in our Go-like model. As seen in Fig. 4, these interactions can both reduce and enhance the resistance of the chain to the mechanical unfolding. This suggests that given the native topology, further optimization with respect to the protein's mechanical stability can be achieved via mutations that alter non-native interactions.

**Acknowledgments**. This work was supported by grants from the Robert A. Welch Foundation and ACS Petroleum Research Fund and by the National Science Foundation CAREER award (to DEM) and by the National Science Foundation grant CSM to GJR.

| $\tilde{v}$ | 0.1 | 1 | 5 | 10 | 15 | 20 | 30 | 50 | 100 | 200 | 500 |
|---|---|---|---|---|---|---|---|---|---|---|---|
| $\tilde{l}_{\Delta W}$ | 0.967 | 0.84 | 0.73 | 0.69 | 0.675 | 0.662 | 0.648 | 0.63 | 0.615 | 0.601 | 0.587 |

Table 1. Single cross link: The dimensionless loop length $\tilde{l}_{\Delta W}$ that maximizes $<\Delta W>$ as a function of the dimensionless pulling velocity $\tilde{v}$.

| $\tilde{v}$ | 0.2 | 1 | 2 | 10 | 20 | 30 | 40 | 60 | 100 |
|---|---|---|---|---|---|---|---|---|---|
| $\tilde{l}_{\Delta W}$ | 0.94 | 0.88 | 0.84 | 0.8 | 0.76 | 0.72 | 0.7 | 0.68 | 0.66 |
| $\tilde{l}_F$ | 0.94 | 0.94 | 0.94 | 0.94 | 0.94 | 0.94 | 0.94 | 0.94 | 0.94 |

Table 2. The NSCL configuration made of three cross-links: The dimensionless loop length $\tilde{l}_{\Delta W}$ that maximizes $<\Delta W>$ and the dimensionless loop length $\tilde{l}_{F_m}$ that maximizes $<F_m>$, as functions of the dimensionless velocity $\tilde{v}$.

**Figure Captions**

Figure 1. Unfolding of a cross-linked chain. (a) The configuration of a $L = 50$ chain with the cross-links $\{\{7,19\},\{15,47\},\{16,42\},\{21,35\},\{40,48\}\}$. (b) The force-vs.-extension curve of this chain in the case of the deterministic unfolding scenario (Model I). Each maximum corresponds to the rupture of one or more cross-links. The mechanical resistance of the chain is characterized by two parameters: The excess work $\Delta W$ required to extend the cross-linked chain relative to that for the "denatured" chain (equal to the shaded area) and the maximum force $F_m$.

Figure 2. An optimal NSCL configuration of an $L=50$ chain with $N=3$ cross-links. Within Model I, this configuration optimizes both $\Delta W$ and $F_m$. In general, the optimal configurations have the form $\{\{i,i+l\},\{i+1,i+2+l\},\{i+3,i+3+l\}\}$ where $l$ is the loop length. For Model II, the loop length $l$ that optimizes $\langle \Delta W \rangle$ is a function of the pulling velocity $v$ while $\langle F_m \rangle$ is optimized by $l = 47$ regardless of the pulling velocity.

Figure 3. (a) The maximum force $\langle F_m \rangle$ and (b) the excess work $\langle \Delta W \rangle$ as a function of the pulling rate for NSCL configurations with different values of the loop length $l$.

Figure 4. Probability distributions for (a) $F_m$ and (b) $\Delta W$ for randomly generated configurations containing $(L+1)/2$ cross-links ($L = 19$), and configurations including the clamp $C_3^*$ with the remaining 7 cross-links generated randomly. The fully random configurations are denoted by the squares and those containing the clamp by the circles.

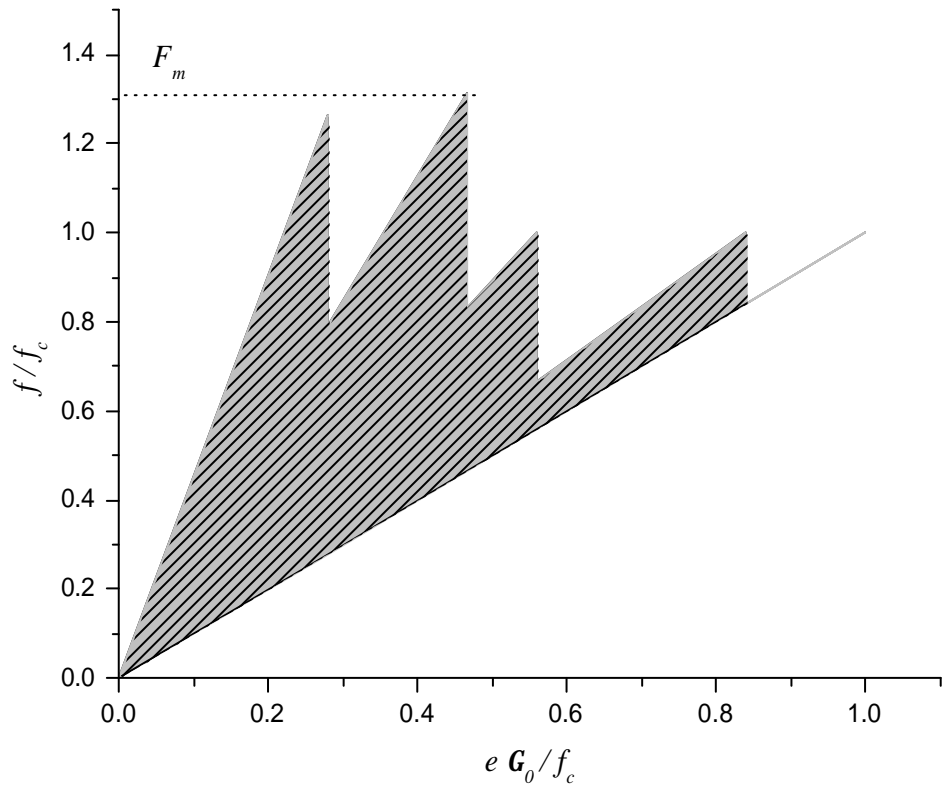

Figure 1

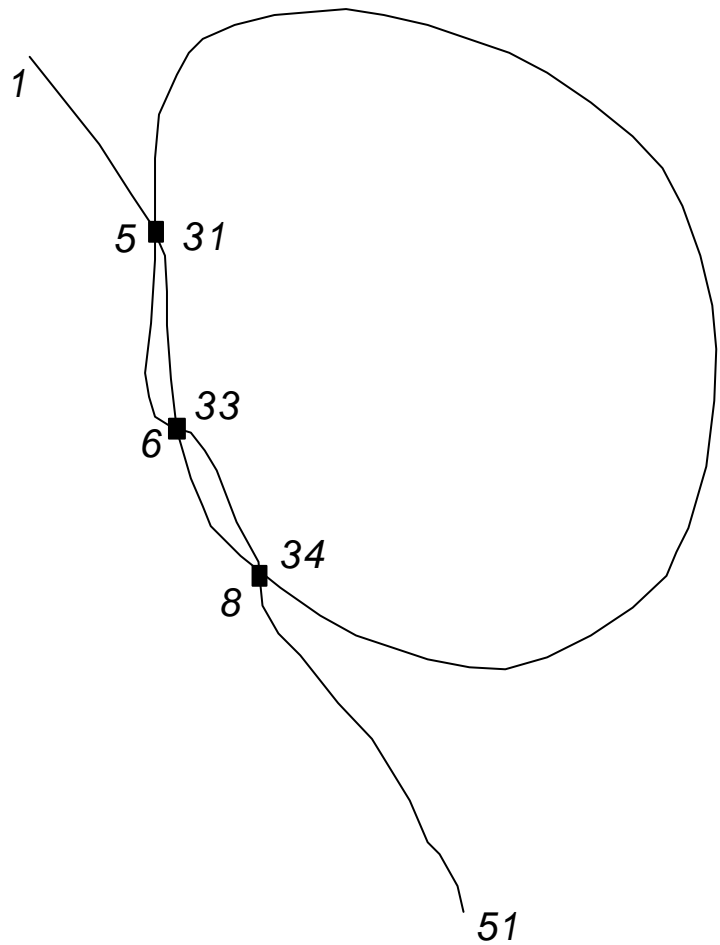

Figure 2

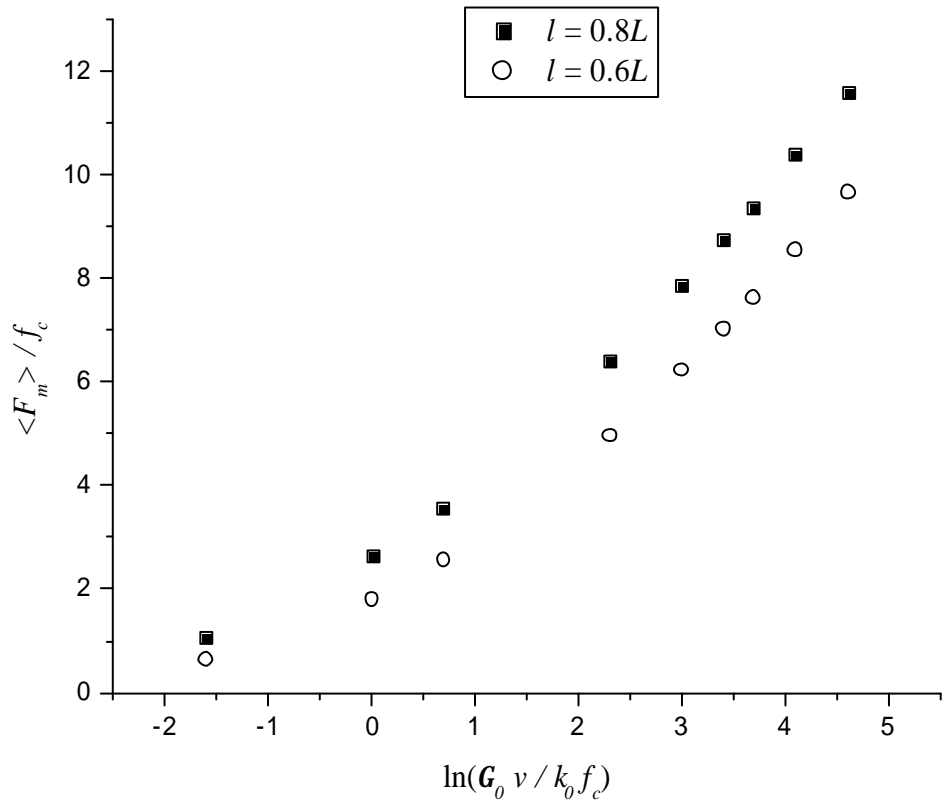

Figure 3a

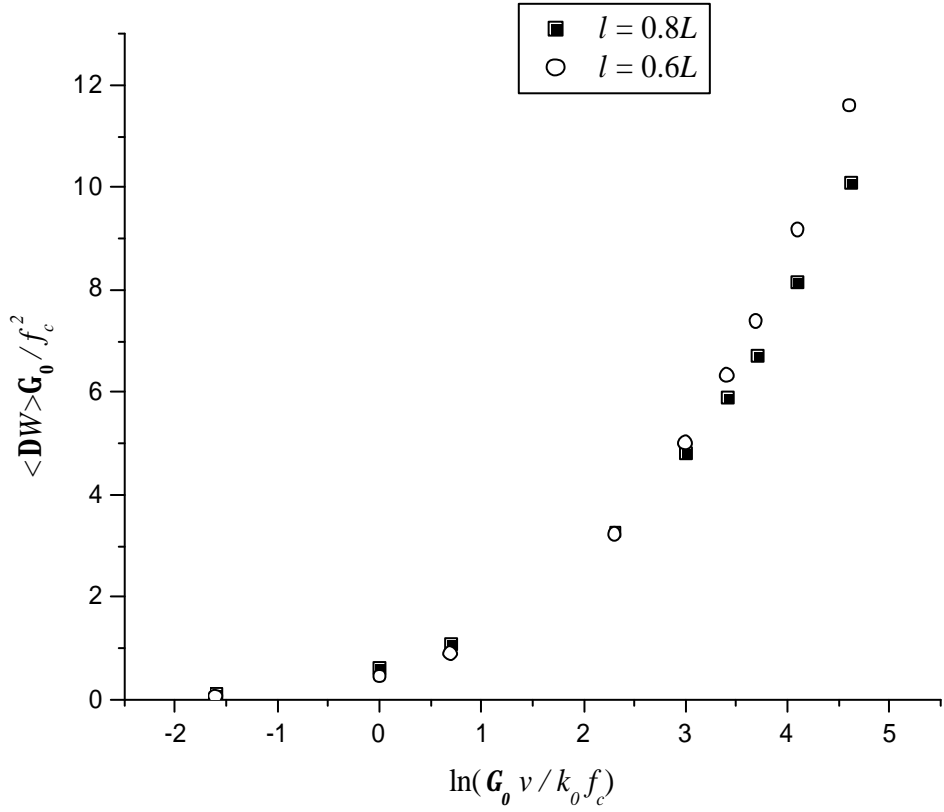

Figure 3b

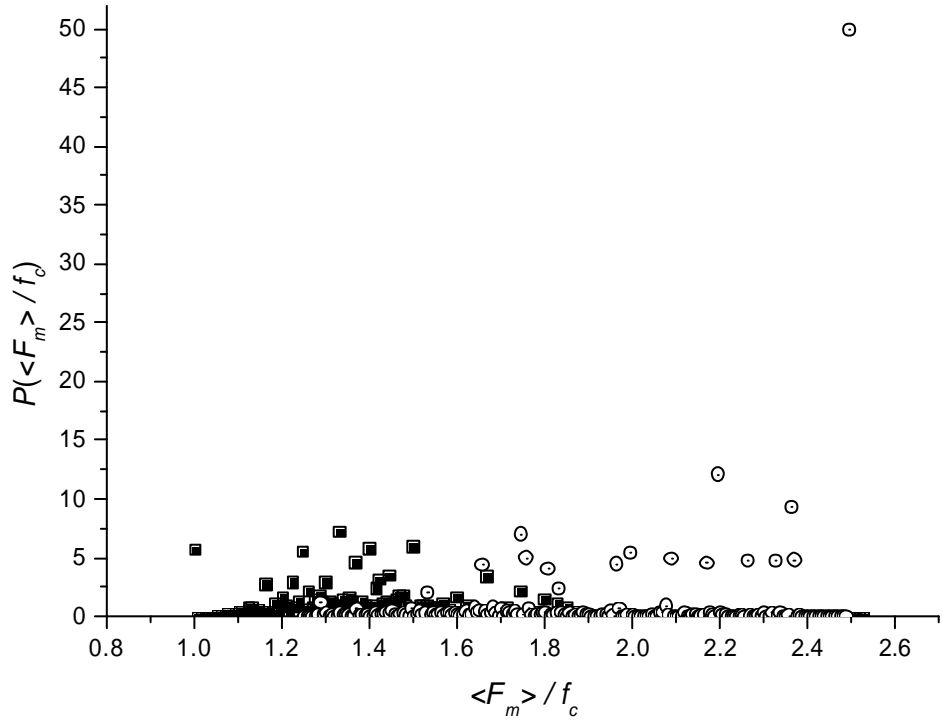

Figure 4a

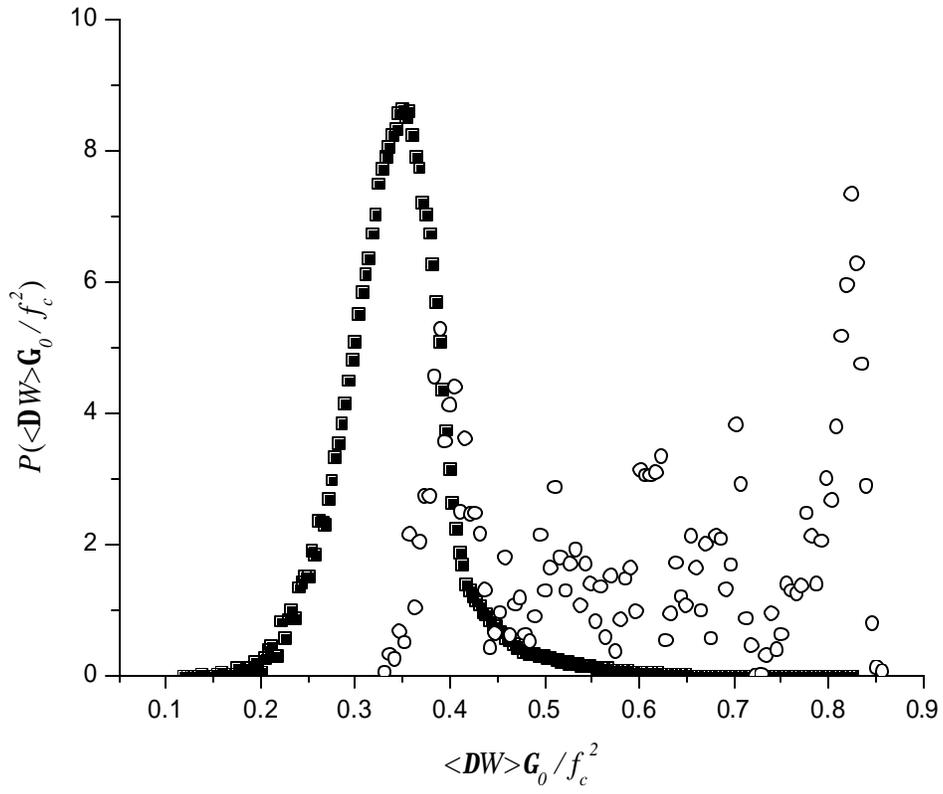

Figure 4b